\documentclass{article}
\usepackage{emulateapj}
\usepackage{amsmath}
\usepackage{amssym}
\usepackage{psfig}
\usepackage{times}

\input{epsf}

\def\dd{{\rm d}}

\def\Uf{U_{\rm free}}

\def\dM{\dot{M}}

\def\sT{\sigma_{\rm T}}

\def\tt{\tilde{t}}

\def\tv{\tilde{v}}

\def\tp{\tilde{p}}

\def\tl{\tilde{\lambda}}

\def\tvrms{\tilde{v}_{\rm rms}}

\def\GCM{\Gamma_{\rm CM}}
\def\bCM{\beta_{\rm CM}}

\def\rin{r_{\rm in}}
\def\tin{t_{\rm in}}

\def\Grms{\Gamma_{\rm rms}}
\def\bG{\Gamma_{\rm av}}

\newbox\grsign \setbox\grsign=\hbox{$>$} \newdimen\grdimen \grdimen=\ht\grsign
\newbox\simlessbox \newbox\simgreatbox \newbox\simpropbox
\setbox\simgreatbox=\hbox{\raise.5ex\hbox{$>$}\llap
     {\lower.5ex\hbox{$\sim$}}}\ht1=\grdimen\dp1=0pt
\setbox\simlessbox=\hbox{\raise.5ex\hbox{$<$}\llap
     {\lower.5ex\hbox{$\sim$}}}\ht2=\grdimen\dp2=0pt
\setbox\simpropbox=\hbox{\raise.5ex\hbox{$\propto$}\llap
     {\lower.5ex\hbox{$\sim$}}}\ht2=\grdimen\dp2=0pt

\def\simlt{\mathrel{\copy\simlessbox}}

\newcommand{\bez}{\begin{eqnarray*}}
\newcommand{\eez}{\end{eqnarray*}}
\newcommand{\be}{\begin{equation}}
\newcommand{\ee}{\end{equation}}
\newcommand{\beq}{\begin{eqnarray}}
\newcommand{\eeq}{\end{eqnarray}}
\newcommand{\bc}{\begin{center}}
\newcommand{\ec}{\end{center}}

\lefthead{BELOBORODOV} 
\righthead{EFFICIENCY OF INTERNAL SHOCKS}

\begin{document}

\title{On the efficiency of internal shocks in gamma-ray bursts}

\author{Andrei M. Beloborodov\altaffilmark{1}} 
\affil{Stockholm Observatory, SE-133 36, Saltsj{\"o}baden, Sweden;
andrei@astro.su.se} 
\altaffiltext{1}{Also at Astro-Space Center of Lebedev Physical Institute, 
Profsojuznaja 84/32, Moscow 117810, Russia}

\begin{abstract}
The fraction of a fireball kinetic kinetic energy that can be radiated away
by internal shocks is sensitive to the amplitude of initial fluctuations
in the fireball. 
We give a simple analytical description for dissipation of modest-amplitude
fluctuations and confirm it with direct numerical simulations. 
At high amplitudes, the dissipation occurs in a non-linear regime
with efficiency approaching 100~\%. Most of the explosion energy is then 
radiated away by the prompt GRB and only a fraction remains to be radiated by 
the afterglow.
\end{abstract}

\keywords{gamma-rays: bursts}

\bigskip

\section{Introduction}

\medskip

Cosmological gamma-ray bursts (GRBs) are huge explosions of energy 
$\sim 10^{53}$~ergs, which may be triggered, e.g., by coalescence of 
neutron stars and/or black holes. The created fireball expands 
relativistically, with a Lorentz factor $\Gamma\sim 10^2$.
In the most likely scenario,
the observed gamma-rays are produced at 
times $t\sim 10^2-10^5$~s after the explosion, when the relativistic outflow
gets optically thin and before it is decelerated by the surrounding medium 
(see, e.g., Piran 1999 for a review). 
The outflow has a fluctuating velocity profile, and 
the $\gamma$-rays are generated by internal shocks that 
develop when faster shells try to overtake slower ones 
(Rees \& M\'esz\'aros 1994).

The internal dissipation was discussed and simulated
numerically in a number of works (e.g., Kobayashi, Sari, \& Piran 1997;
Daigne \& Mochkovitch 1998; Lazzati, Ghisellini, \& Celotti 1999; 
Panaitescu, Spada, \& M\'esz\'aros 1999; Spada, Panaitescu, M\'esz\'aros 1999;
Kumar 1999). It was concluded that only a few percent of the total kinetic 
energy is emitted to infinity (e.g., Panaitescu et al. 1999; Kumar 1999). 
Then the energy of the prompt GRB should be dominated by the early afterglow
associated with an external shock. Observations show the opposite.
Besides, the low efficiency requires huge explosion energies.

The conclusion that the internal dissipation has a low efficiency was based on 
certain assumptions: 
\medskip

\noindent
(1) 
A specific probability distribution was assumed for the outflow Lorentz factor,
such that the root-mean-square (rms) of the fluctuations was less than 
$1/\sqrt{3}$ (see eq. [11]). In fact, the rms can be much higher.
\medskip

\noindent
(2) 
The fluctuations were taken with a white (flat) spectrum.
\medskip

\noindent
(3) 
It was often assumed that the dissipated energy is equally 
distributed between the electrons, protons, and magnetic field, and only the
electron part is radiated. In fact, the radiative 
capability of shocked matter is highly uncertain and dependent on
poorly understood plasma processes.
The high radiative capability is favored by the 
observed high luminosities of the prompt GRBs.
If most of the dissipated energy is transferred to the electrons 
and radiated, the outflow stays at low pressure and the coasting 
matter gets concentrated into thin shells (caustics).
\medskip

In \S~2, we give a simple analytical description for  
dissipation of white fluctuations with modest rms (the linear regime). 
Then the efficiency is proportional to the mean square of the 
initial fluctuations. In \S 3, we study high-amplitude fluctuations.
The efficiency then approaches 100~\%.
The analytical results are illustrated with numerical simulations.


\section{Low-amplitude fluctuations}

The size of the central engine, $\rin\sim 10^7-10^8$~cm corresponds to a 
typical time scale $\tin=\rin/c\sim 1$~ms. The outflow generated by the engine 
can have highly correlated parameters on this time scale, so that portions 
$\Delta t\simlt \tin$ can be considered as individual shells in the continuous
outflow. The outflow is then discretized as a sequence of such shells, 
$i=1,...,N$, with mass $m_i$ and energy input $e_i$. 
One after another, the shells are accelerated to a Lorentz factor 
$\Gamma_i=e_i/m_i$ ($e_i$ gets converted into the bulk kinetic energy)
on the time-scale $\Gamma_i \tin<1$~s (see Piran 1999). Subsequent
evolution of the outflow proceeds in the coasting regime until 
the shells begin to collide. 

Suppose $\Gamma_i$ is a white noise with the average value $\bG$ and initial
rms ${\Grms}_0<\bG$. Masses of the shells can be taken equal, 
$m_i=m_0$, or also fluctuating -- the results will be the same. 
The initial scale of the fluctuations is $\lambda_0\sim\rin$. 
The collision process starts at
\begin{equation}
t_0\sim\frac{\bG^3}{{\Grms}_0}\frac{\lambda_0}{c}.
\end{equation}
After time $t\sim 2t_0$, the first generation of shell coalescence has been 
done (the typical mass of a shell increases by a factor of two), then
the second generation occurs, and so on. In the process of hierarchical
coalescence, the scale of the fluctuations increases, $\lambda=\lambda(t)$, 
as well as the average mass of an individual shell, 
$\lambda(t)/\lambda_0=m(t)/m_0\equiv K(t)$.

The fluctuations look especially simple when viewed
from the frame moving with Lorentz factor $\bG$. We will denote 
quantities measured in this frame by symbols with tilde.
The fluctuation velocity is
$\tv_i/c=(\Gamma_i^2-\bG^2)/(\Gamma_i^2+\bG^2)$.
Given $\Grms<\bG$, we have $\tv_i/c\approx(\Gamma_i-\bG)/\bG$
with the average $\tv_{\rm av}=0$ and the rms,
\begin{equation}
   \frac{\tvrms}{c}\approx \frac{\Grms}{\bG}.
\end{equation}
The outflow motion is thus decomposed into two parts: the relativistic
motion of the center-of-momentum (CM) with a Lorentz factor $\GCM\approx\bG$
and superimposed non-relativistic fluctuations.

The hierarchical coalescence of shells can be expected to occur in a 
self-similar regime, so that the collisional time-scale for one generation
is about the time passed since the explosion,
\begin{equation}
  \frac{\tl}{\tvrms}\approx \tt=\frac{t}{\GCM}.
\end{equation}
Here, $\tl=\GCM\lambda$ is the scale of the fluctuations in the CM-frame.
After coalescence of $K$ shells with initial momentum 
$\tp_0\sim m_0\tv_{{\rm rms}0}$, the new big shell has a momentum 
$\tp\sim\sqrt{K}\tp_0$ (this is the random walk formula: $K$ momenta $\sim p_0$
are summed with random sign). We thus have,
\begin{equation}
  m\tvrms\sim\sqrt{K}m_0\tv_{{\rm rms}0}
  =\sqrt{\frac{m}{m_0}}m_0\tv_{{\rm rms}0}.
\end{equation}
Combining (2), (3) and (4), we get the self-similar solution describing the 
hierarchical coalescence at $t>t_0$,
\begin{equation}
  \frac{m}{m_0}=\frac{\lambda}{\lambda_0}=\left(\frac{t}{t_0}\right)^{2/3}, 
  \quad \frac{\Grms}{{\Grms}_0}=\left(\frac{t}{t_0}\right)^{-1/3}. 
\end{equation}
Note that the collisions establish
a Gaussian distribution of $\tv$. One can therefore prescribe 
a temperature to the system of shells, $T$, which is related 
to the average kinetic energy by $m\tvrms^2/2=kT/2$ 
(shells have one degree of freedom). 
From equation (5) it follows that $T=const$.
The evolution of shells can thus be described as isothermal sticking together. 

The free energy of the outflow is given by 
\begin{equation}
\Uf=\GCM M\frac{\tvrms^2}{2}\approx Mc^2\frac{\Grms^2}{2\GCM},
\end{equation}
where $M$ is the total mass of the outflow. This energy is available for 
dissipation. From equation (5) one gets 
\begin{equation}
   \Uf=\Uf^0 \left(\frac{t}{t_0}\right)^{-2/3}. 
\end{equation}
In Figure~1, we illustrate solution (5) by direct numerical 
simulations. 

Most of the free energy is dissipated at $t\sim t_0$. It should be compared 
with the time at which the outflow gets optically thin, 
\begin{equation}
t_*\approx\frac{\dM\sT}{8\pi m_pc^2\Gamma^2}
  =\frac{L\sT}{8\pi m_pc^4\Gamma^3}\approx 2\times 10^2 L_{52}\Gamma_2^{-3},
\end{equation}
where $\dM$ is the mass outflow rate and $L=\dM c^2\bG$ is the kinetic 
luminosity (assuming a spherical outflow). In the case of 
$\GCM<\Gamma_*\approx 180(\Grms/\GCM)^{1/5}L_{52}^{-1/5}
(\lambda_0/3\times 10^7)^{-1/5}$, dissipation starts at $t_0<t_*$. 
The free energy remaining in the outflow by
the transparency moment is $\Uf^*=\Uf^0(t_*/t_0)^{-2/3}$, and most of the 
energy is dissipated at the optically thick stage. 
The radiation produced at $t<t_*$ is trapped in the 
plasma whose volume increases proportionally to $t^2$.
As a result of adiabatic cooling, the volume-integrated radiation energy is 
reduced as $t^{-2/3}$. The energy conservation law implies that 
the adiabatic cooling is accompanied by a regular radial acceleration of 
shells in the outward direction. The energy of trapped radiation is thus spent 
to accelerate the outflow CM, and it is lost as a free energy. 
A fraction $(t_*/t)^{-2/3}$ of the radiation energy survives till $t_*$ and 
contributes to the observed luminosity, 
$$
L_*\approx\int_{t_0}^{t_*}\left(-\frac{\dd\Uf}{\dd t}\right)
    \left(\frac{t_*}{t}\right)^{-2/3}\dd t +\Uf^*.
$$
The radiative efficiency (the ratio of $L_*$ to the explosion energy) is
\begin{equation}
 \epsilon\approx\frac{L_*}{\GCM M c^2} \approx 
   \frac{A^2}{2}
 \left(\frac{t_*}{t_0}\right)^{-2/3}\left(\frac{2}{3}\ln\frac{t_*}{t_0}
       +1\right). 
\end{equation}
Here, $A={\Grms}_0/\GCM$ is the initial amplitude of the fluctuations.

In the case of $\GCM>\Gamma_*$, we have $t_0>t_*$ and all the radiated
free energy will escape. Then, 
\begin{equation}
  \epsilon\approx\frac{\Uf^0}{\GCM Mc^2}=\frac{A^2}{2}.
\end{equation}

If the initial Lorentz factor takes
random values between $\Gamma_{\rm min}$ and $\Gamma_{\rm max}$
(as assumed in most of the previous works), then 
\begin{equation}
   A^2=\frac{1}{3}-\frac{4\psi}{3(1+\psi)^2}, \qquad 
   \psi\equiv\frac{\Gamma_{\rm min}}{\Gamma_{\rm max}}.
\end{equation}

Since $A^2<1/3$ in equation (11), $\epsilon$ does not exceed $\approx 15$~\%.  
If one further assumes that only 1/3 of the 
energy is actually radiated and the other 2/3 are 
stored as internal energy subject to adiabatic cooling, then the efficiency
is reduced by a factor of 3 (2/3 of the dissipated energy is then
spent to accelerate the CM). One thus arrives at $\sim 5$~\% limit on 
$\epsilon$.

\centerline{
\epsfxsize=8.1cm {\epsfbox{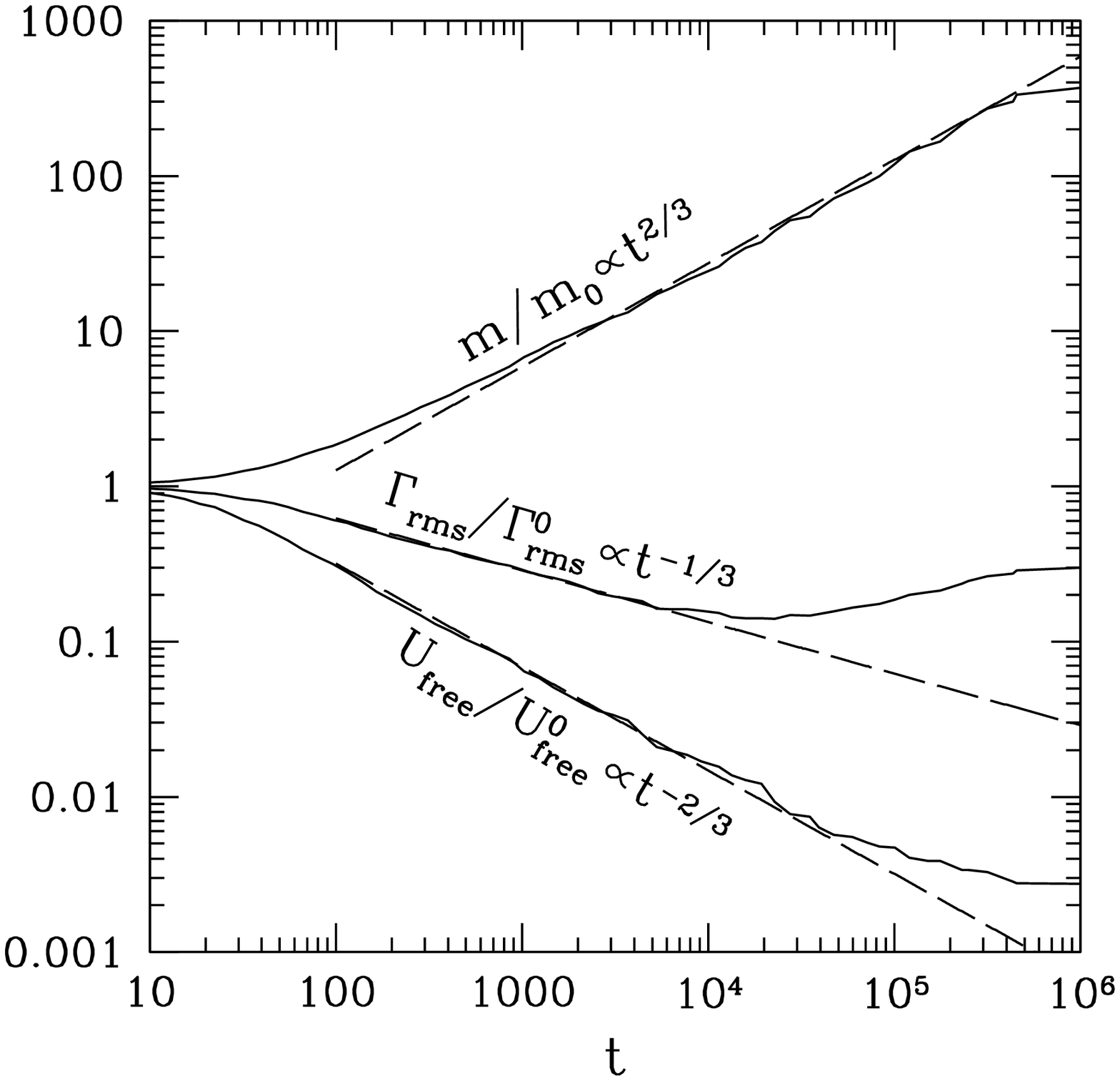}}
}
\figcaption{ Evolution of $N=3000$ shells with initial separation 1~ms and
$\Gamma$ fluctuating according to equation (12) with $A=0.2$ and 
$\Gamma_0=100$. 
The shells begin to collide at $t_0\sim 50$~s. {\it Solid curves} 
show the average mass of the merging shells, $m(t)$, the rms of their Lorentz
factors, $\Grms(t)$, and the free energy of the system, $\Uf(t)$.
{\it Dashed lines} display the self-similar solution (5,7). 
The deviations from the power-laws occur at late times when the number of
shells is reduced and most of the free energy has been dissipated.
\label{fig1}}

Numerical simulations of Kobayashi et al. (1997; see their Table~1) are in 
good agreement with equations (10,11), except for the case where
$\psi\rightarrow 0$ and $m_i\propto \Gamma_i^{-1}$ (their parameter $\eta=-1$).
One can show that in this special case 
$\GCM/\Grms\approx \sqrt{\psi}\rightarrow 0$
and $\epsilon\approx 1+\sqrt{\psi}\ln\psi\rightarrow 1$. 
The outflow dynamics is then non-linear (even though $A=1/\sqrt{3}<1$),
similar to the high amplitude case we study in \S~3.


\section{High-amplitude fluctuations}

Suppose that $\Gamma-1$ fluctuates with a log-normal distribution,
\begin{equation}
   \ln\frac{\Gamma-1}{\Gamma_0-1}=A\xi, \qquad 
   P(\xi)=\frac{e^{-\xi^2/2}}{\sqrt{2\pi}}.
\end{equation}
Here, $A$ measures the amplitude of the fluctuations. At $A<1$, we have 
$(\Gamma-\Gamma_0)/\Gamma_0\approx A\xi$, so that $\bG=\Gamma_0$ and 
$\Grms/\bG=A$. At $A>1$ we are in the high-amplitude regime.

The possibility of high efficiencies at $A>1$
can be understood when comparing two global characteristics of the
outflow: the CM Lorentz factor and the specific energy, $\eta$
(in units of $c^2$). For definiteness, take a uniform initial distribution of 
mass. Then,
\begin{equation}
  \bCM=\frac{\int P(\xi)\Gamma\beta \dd \xi}
            {\int P(\xi)\Gamma\dd \xi}, 
             \quad \GCM=\frac{1}{\sqrt{1-\bCM^2}},
\end{equation}
\begin{equation}
 \quad \eta=\int P(\xi)(\Gamma-1)\dd \xi.
\end{equation}
It easy to show that at high amplitudes $\eta\gg\GCM$, i.e., the chaotic (free) 
component
of the kinetic energy is much larger than the regular component. If the 
dissipation occurs in the optically thin regime and the emitted radiation 
is approximately isotropic in the CM-frame, then the momentum conservation law 
implies $\GCM\approx const$. When the dissipation is done, the final kinetic 
energy is about $(\GCM-1)Mc^2\ll\eta Mc^2$, i.e., almost all the explosion 
energy has been radiated to infinity. 

In fact, in the high-amplitude case, the fluctuations start to dissipate 
very early, much before the transparency moment.
The optically thick dissipation accelerates the CM, as discussed
in \S~2: the free energy gets transformed into the bulk motion of the outflow 
as a whole. The process is non-linear: the CM is substantially accelerated,
in contrast to the low-amplitude case. It tends to reduce the 
initially big difference between $\GCM$ and $\eta$ by increasing $\GCM$,
while $\eta$ stays at the initial value, $\eta=\eta_0$.

If at the transparency moment the ratio $\eta/\GCM$ is still high, then
a high efficiency can be expected: the dissipation at 
the optically thin stage converts the difference between $\eta Mc^2$
and $(\GCM-1)Mc^2$ into the observed radiation. Note also, that the radiation 
produced before the transparency moment is not completely lost. Even being
cooled adiabatically, it contributes substantially to the outgoing luminosity. 

We now illustrate with numerical simulations. We generate a sequence of
$3\times 10^3$ thin shells with initial separation 1~ms.
Their Lorentz factor fluctuates according to equation (12). 
The shells have equal mass $m_0$. Note that the results depend on the initial 
mass distribution (in contrast to the linear case),
and $m_i=m_0$ is taken as a simple example. 
The duration of the central engine activity is 3~s.
The results do not change substantially if one takes longer activity:
each 1~s portion of the outflow is causally disconnected from the other 
portions during the main emission time, $t<10^4$~s.  

The transparency moment, $t_*$, is roughly estimated by equation (8) with 
$\Gamma\sim\GCM$.
The transition from the optically thick to optically thin regime is treated
in the simplest way: (1) If two shells merge at $t<t_*$, we assume that no 
radiation is emitted. Instead, the radiation is trapped and contributes
to the kinetic energy of the new big shell. In other words, shell coalescence 
at $t<t_*$ proceeds with conservation of energy, $\eta=const$. A fraction 
$(t_*/t)^{-2/3}$ of radiation trapped at moment $t$ 
survives till $t_*$ and contributes to the luminosity, $L_*$.
(2) At $t>t_*$, only specific momentum conserves in the coalescence events. 
The energy released in the inelastic collision is radiated away 
isotropically in the rest frame of the newly formed shell.

We consider models of two types. Model I: $t_*=200$~s and
$\Gamma_0$ in equation (12) is adjusted in such a way that  
$\GCM\approx 100$ after the dissipation is finished. Model II:
$t_*=7$~s and $\Gamma_0$ is adjusted to get the final $\GCM\approx 300$. 
In both models, the isotropic kinetic luminosity, $L\sim 10^{52}$~erg~s$^{-1}$. 

The outflow forms by $t=3$~s and after that it has a well defined CM with 
a velocity $\bCM=Pc/E$, where $P$ is the total momentum and $E=(\eta+1)Mc^2$ 
is the total energy of the outflow. 
(Note that the initial $\eta=\eta_0$ depends on the specific statistical 
realization $\Gamma_i$ and it varies around the average expected 
value given by eq.~[14]).  
In Figure~2, we show examples of the non-linear evolution of $\eta$ and $\GCM$
in Models I and II. At high $A$, the radiation is mostly produced by shells 
moving faster than the CM, and this results in the CM deceleration at $t>t_*$.

Then, we compute a sequence of models and find the 
efficiency, $\epsilon\equiv L_*/\eta_0 Mc^2$, as a function of $A$ (Fig.~3).
At $A<1$, we are in the linear regime of \S~2. Shells begin to collide at 
$t_0\approx A^{-1}\Gamma_0^2(\lambda_0/c)$ and then evolve according to 
solution (5). Model~II ($\GCM\approx 300>\Gamma_*$) is in 
perfect agreement with equation (10).
The efficiency of Model~I ($\GCM\approx 100<\Gamma_*$) is reduced
as a result of adiabatic losses during the optically
thick stage. 

\centerline{
\epsfxsize=9.0cm {\epsfbox{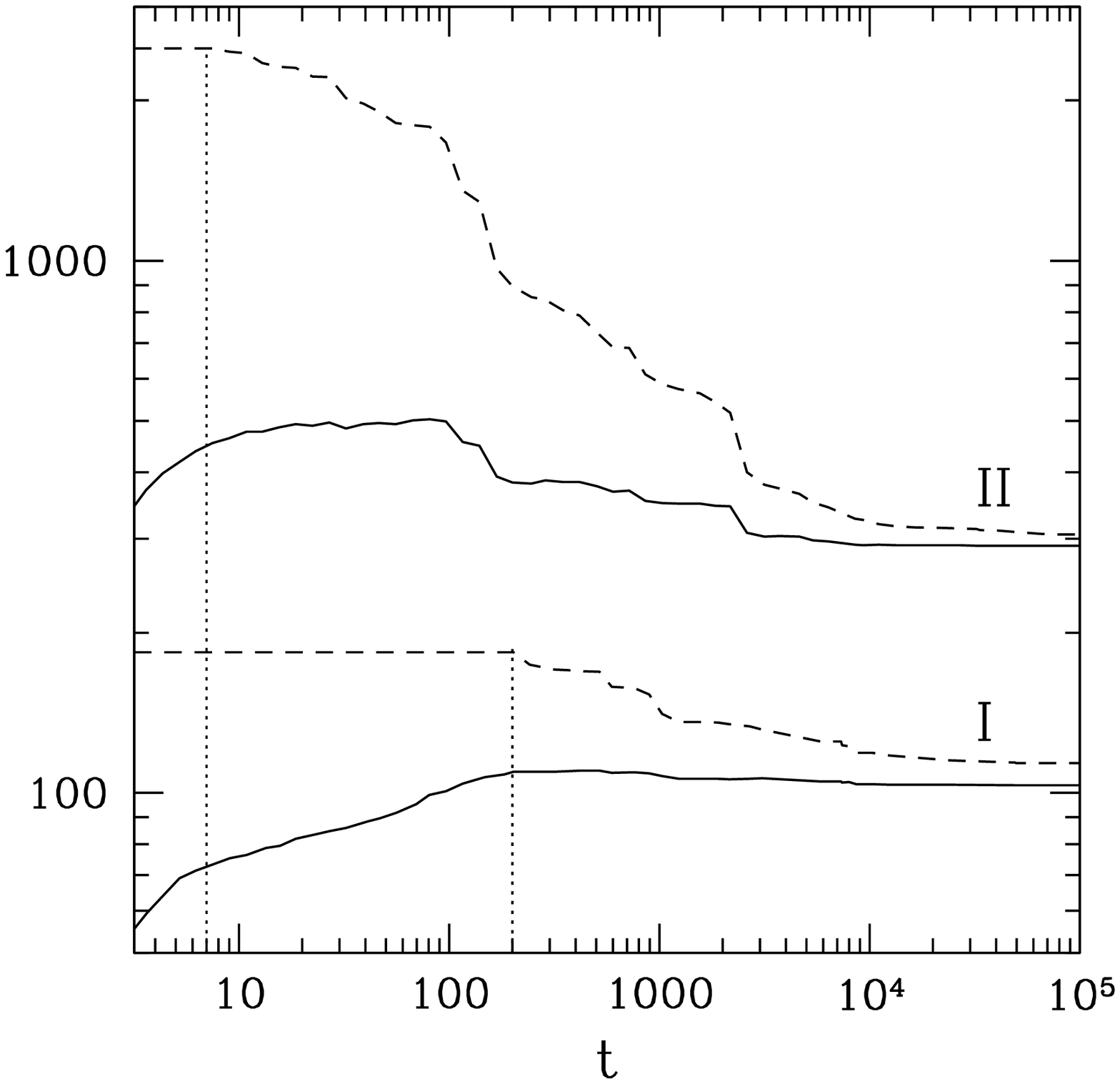}}
}
\figcaption{ 
Examples of fireball evolution with a strongly fluctuating 
Lorentz factor ($A=4$ in eq.~[12]) for Model I and II (see the text). 
The outflow is formed during the first 3~s.
{\it Dashed curves} show the specific energy of the outflow, $\eta(t)$,
and {\it solid curves} show the center-of-momentum Lorentz factor, $\GCM$.
{\it Dotted lines} mark the transparency moments, $t_*$, for the two models.
\label{fig1}}

At $A>1$, the efficiency increases up to $\sim 83$~\% in Model~I, 
and up to 96~\% in Model~II.
The global parameters of the outflow show substantial 
variations depending on the particular realization of the initial 
$\Gamma_i$. 
Figure~3 shows $\epsilon$ averaged for many realizations and its
standard deviation.

Figure~4 shows examples of the generated light curves. Each coalescence event 
produces a pulse of a standard shape corresponding to a thin instantaneously
radiating shell. The observed pulse has a width 
$\Delta t_{\rm obs}\sim t/\Gamma^2$, where $t$ is the time at 
which the radiation leaves the shell.
Although the initial fluctuations are Gaussian, the non-linear dissipation 
generates a highly correlated signal which is a mixture of quiescent periods 
and periods of strong activity. Such a behavior is observed in GRBs and this
special feature is naturally explained by the non-linear model. 

There is another special feature of the non-linear regime.
The effective temperature of fluctuations in the CM-frame is relativistic,
and an additional emission mechanism appears: inverse Compton 
scattering (IC) on the bulk motions (see also Lazzati et al. 1999).
This mechanism does not work at the optically thick stage since the radiation
is trapped and follows the motions of shells. However, when the outflow gets
optically thin, the radiation can propagate and promote
momentum exchange between the shells without direct collisions (like Silk 
damping in the early Universe). This effect should be accounted for in future.
We expect that it will not change crucially the efficiency because: 
(1) Anyway, the free energy is radiated away,
whatever dissipation mechanism works.
(2) Photon exchange does not crucially enhance the rate of dissipation because
direct collisions also occur with relativistic velocities in the CM-frame.

The IC by bulk motions can provide an observational test for the model.
In particular, it can explain the high energy tail 
sometimes observed in GRB spectra. The photons emitted by internal 
shocks (with energy $\sim 10-1000$~keV) are boosted in energy by a factor of 
$(\Grms/\GCM)^2$, resulting in GeV emission.

The study of fluctuating fireballs in this paper was limited to the 
case of white fluctuations. Fluctuations with 
an arbitrary spectrum will be studied in a future paper (Beloborodov 2000).

\acknowledgments

This work was supported by the Swedish Natural Science Research Council and 
RFBR grant 00-02-16135.

\centerline{
\epsfxsize=9.0cm {\epsfbox{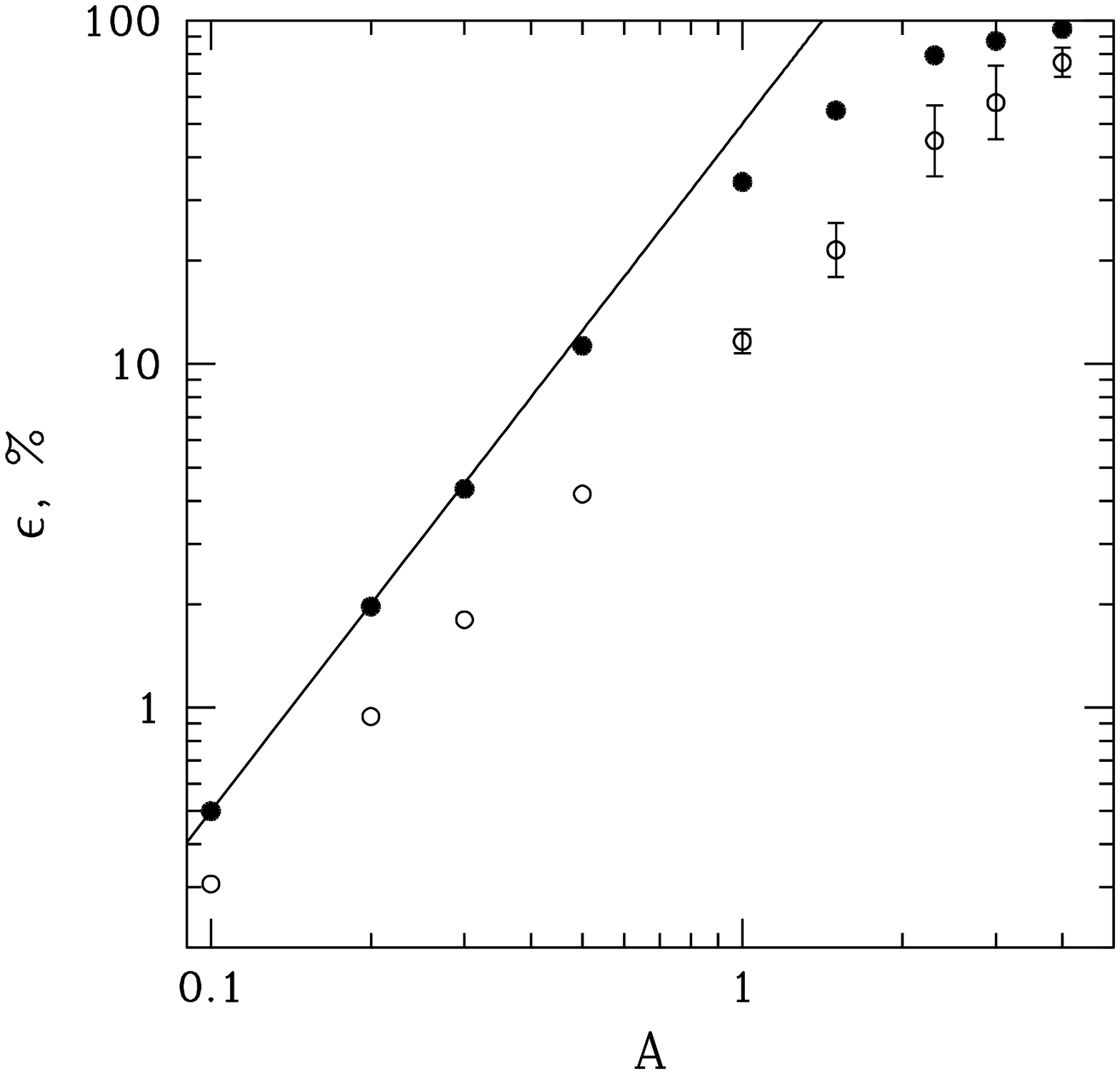}}
}
\figcaption{ 
Efficiency, $\epsilon$, as a function of the fluctuation amplitude, $A$.
{\it Open circles} show the results in Model~I (with final $\GCM\approx 100$) 
and {\it filled circles} show Model~II (with final $\GCM\approx300$).
{\it Bars} show the standard deviation (where it is larger than the symbol 
size).
{\it Line} displays $\epsilon=A^2/2$ expected in the optically thin linear 
regime (see eq.~[10]).  
\label{fig1}}

\centerline{
\epsfxsize=9.0cm {\epsfbox{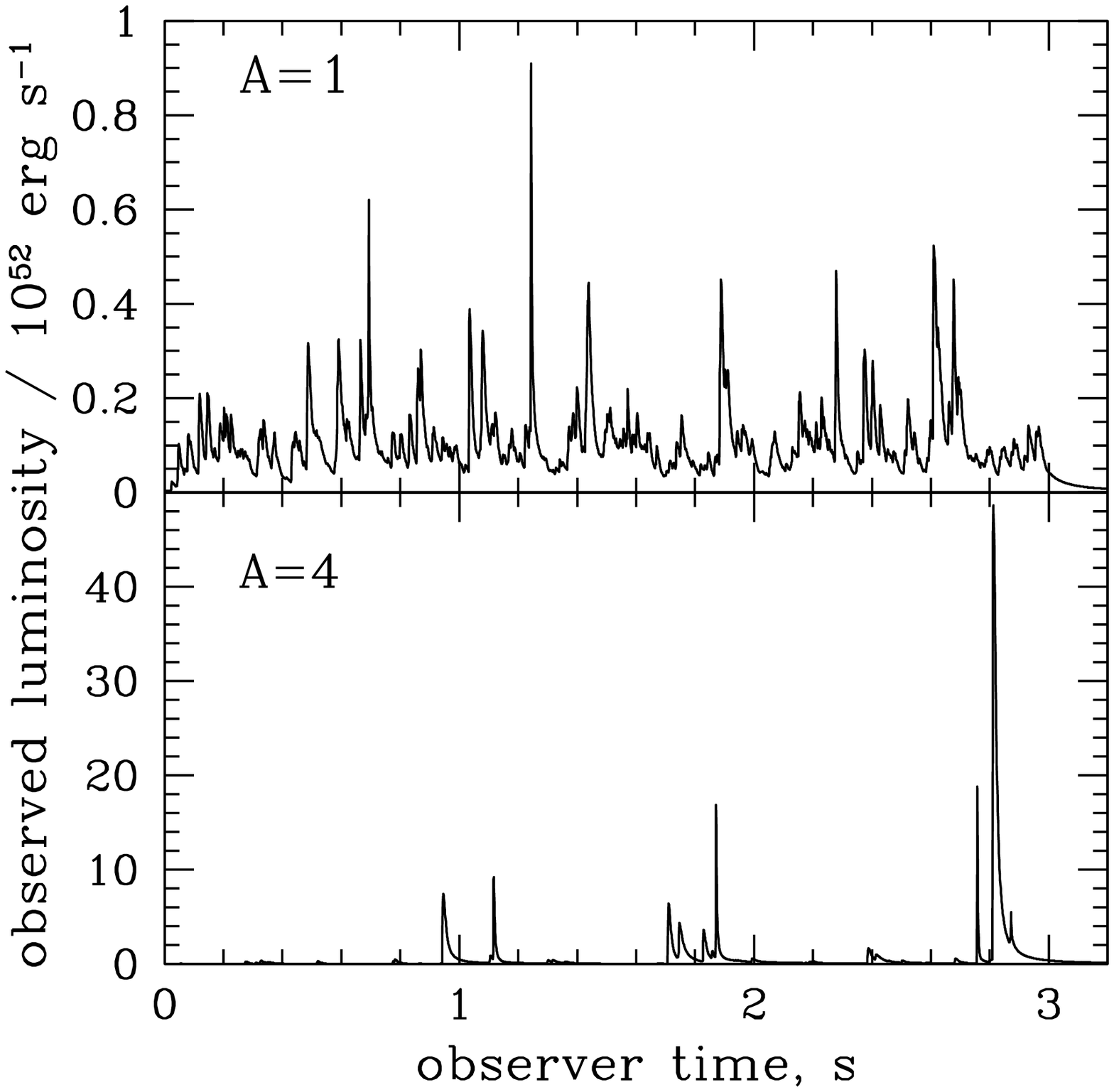}}
}
\figcaption{ 
Example light curves in Model~I with $A=1$ and $A=4$.
\label{fig1}}

\end{document}